\documentclass[10pt,aps,prb,twocolumn,preprintnumbers,amsmath,amssymb,floatfix,showpacs,citeautoscript,superscriptaddress]{revtex4-2}
\usepackage[utf8]{inputenc}
\usepackage[T1]{fontenc}
\usepackage[english]{babel}
\usepackage[pdftex]{graphicx}
\usepackage[normalem]{ulem}
\usepackage{multirow}
\usepackage{amsmath}
\usepackage{times}
\usepackage{color}
\usepackage[usenames,dvipsnames,svgnames,table]{xcolor}
\usepackage[colorlinks,plainpages=false,linkcolor=blue,urlcolor=blue,citecolor=blue,pdfpagemode=UseNone]{hyperref}
\usepackage{svg}
\usepackage{notes2bib}
\usepackage{siunitx}
\usepackage{chemformula}

\renewcommand{\AA}{\text{\r{A}}}
\newcommand{\mrm}[1]{\mathrm{#1}}
\newcommand{\mbf}[1]{\mathbf{#1}}
\newcommand{\Tc}{T_{\mrm{c}}}
\newcommand{\TcAD}{T_{\mrm{c}}^{\mrm{AD}}}

\newcommand{\Eh}{E_{\mrm{h}}}
\newcommand{\Ef}{E_{\mathrm{f}}}
\newcommand{\Eg}{E_{\mathrm{g}}}

\newcommand{\wtwo}{\omega_{2}}

\newcommand{\atf}{\alpha^2F(\omega)}

\newcommand{\bk}{\mbf{k}}
\newcommand{\bq}{\mbf{q}}

\definecolor{d12orange}{rgb}{0.8500    0.3250    0.0980}
\definecolor{g8yellow}{rgb}{0.    0.6    0.298}
\definecolor{ppurple}{rgb}{0.4940    0.1840    0.5560}

\definecolor{mag}{RGB}{255,0,255}

\begin{document}

\title{Developing a Complete AI-Accelerated Workflow for Superconductor Discovery}

\author{Jason B. Gibson}
\email{jasongibson@ufl.edu}
\affiliation{Quantum Formatics, Cambridge, Massachusetts 02139, USA}
\affiliation{Department of Materials Science and Engineering, University of Florida, Gainesville, Florida 32611, USA}
\affiliation{Quantum Theory Project, University of Florida, Gainesville, Florida 32611, USA}

\author{Ajinkya C. Hire}
\affiliation{Department of Materials Science and Engineering, The Pennsylvania State University, University Park, Pennsylvania 16802, USA}
\affiliation{Department of Materials Science and Engineering, University of Florida, Gainesville, Florida 32611, USA}
\affiliation{Quantum Theory Project, University of Florida, Gainesville, Florida 32611, USA}

\author{Pawan Prakash}
\affiliation{Quantum Theory Project, University of Florida, Gainesville, Florida 32611, USA}
\affiliation{Department of Physics, University of Florida, Gainesville, Florida 32611, USA}

\author{Philip M. Dee}
\affiliation{Computational Sciences and Engineering Division, Oak Ridge National Laboratory, Oak Ridge, Tennessee 37831, USA}

\author{Benjamin Geisler}
\affiliation{Department of Materials Science and Engineering, University of Florida, Gainesville, Florida 32611, USA}
\affiliation{Department of Physics, University of Florida, Gainesville, Florida 32611, USA}

\author{Jung Soo Kim}
\affiliation{Department of Physics, University of Florida, Gainesville, Florida 32611, USA}
\author{Zhongwei Li}
\affiliation{Department of Physics, University of Florida, Gainesville, Florida 32611, USA}
\author{James J. Hamlin}
\affiliation{Department of Physics, University of Florida, Gainesville, Florida 32611, USA}
\author{Gregory R. Stewart}
\affiliation{Department of Physics, University of Florida, Gainesville, Florida 32611, USA}

\author{P. J. Hirschfeld}
\affiliation{Department of Physics, University of Florida, Gainesville, Florida 32611, USA}
\author{Richard G. Hennig}
\affiliation{Department of Materials Science and Engineering, University of Florida, Gainesville, Florida 32611, USA}
\affiliation{Quantum Theory Project, University of Florida, Gainesville, Florida 32611, USA}
\affiliation{Department of Physics, University of Florida, Gainesville, Florida 32611, USA}

\date{\today}

\begin{abstract}
The quest to identify new superconducting materials with enhanced properties is hindered by the prohibitive cost of computing electron-phonon spectral functions, severely limiting the materials space that can be explored. Here, we introduce a Bootstrapped Ensemble of Equivariant Graph Neural Networks (BEE-NET), a machine-learning model trained to predict the Eliashberg spectral function and superconducting critical temperature with a mean-absolute-error of 0.87~K relative to DFT-based Allen-Dynes calculations. Intriguingly, BEE-NET achieves a true-negative-rate of 99.4\%, enabling highly efficient screening for the rare property of superconductivity. Integrated into a multi-stage, AI-accelerated discovery pipeline that incorporates elemental-substitution strategies and machine-learned interatomic potentials, our workflow reduced over 1.3 million candidate structures to 741 dynamically and thermodynamically stable compounds with DFT-confirmed $T_{\mathrm{c}} > 5$~K. We report the successful synthesis and experimental confirmation of superconductivity in two of these previously unreported compounds. This study establishes a data-driven framework that integrates machine learning, quantum calculations, and experiments to systematically accelerate superconductor discovery.
\end{abstract}

\maketitle

\section{Introduction}

Theories of superconductivity, from Bardeen-Cooper-Schrieffer (BCS) to Migdal-Eliashberg extensions~\cite{Bardeen1957, Eliashberg1960, Eliashberg1961}, have successfully explained the properties of conventional superconductors, and significant progress has been made in unraveling the mechanisms behind unconventional superconductivity~\cite{Keimer2015,Fernandes2022}. Despite these advances, theory has historically played a limited role in guiding the discovery of new superconducting materials. Over the past decade, however, advances in density functional theory (DFT) for superconductivity~\cite{Boeri2020,Pickett2023} and the rise of machine learning (ML)~\bibnote{See S. Xie et al, ``Towards high-throughput superconductor discovery via machine learning'' in Ref.~\cite{ROADMAP}} have challenged Bernd Matthias’ infamous dictum, ``avoid theorists,''~\cite{Mazin2010} as a guiding principle in the search for superconductors. First-principles calculations of the superconducting critical temperature $\Tc$ for electron-phonon-driven systems, long deemed infeasible, are now routine for small unit cells, yielding results with useful accuracy~\cite{Giustino2017, Oliveira88, LuedersSCDFTI2005, MarquesSCDFTIIMetals2005, Pickett2023, Marzari_dft_superC}.
Despite this progress, the computational cost remains high, rendering large-scale screening of materials databases for new superconductors impractical.
Machine learning offers a low-cost alternative to \textit{ab initio}-based scans of large databases and a potential avenue for discovering entirely new superconductors. 
However, efforts to develop ML models for rapidly and accurately predicting superconducting properties have been hindered by a lack of suitable training databases. Unlike other materials discovery problems, the pool of known electron-phonon superconductors is relatively small, numbering only in the thousands.  Some studies~\cite{Zeng2019, Li2020, Roter2020, Konno2021, Kim_Dordevic2023, Stanev2018} have taken advantage of the largest available resource, the SuperCon database~\cite{supercon2011}, which contains experimental $\Tc$ values for approximately 30,000 materials. However, this database contains many unconventional superconductors, as well as erroneous records~\cite{Stanev2018, Hamidieh2018, Kim_Dordevic2023, Sommer2022_3DSC}, leading some groups to construct smaller, more carefully curated alternative databases~\cite{Hosono2015, Sommer2022_3DSC, Foppiano2023}. Others, including the current authors, have instead focused on generating computational databases of the Eliashberg spectral function, $\alpha^2F(\omega)$, for thousands of materials calculated directly from DFT, from which $\Tc$ and the superconducting gap function can be determined~\cite{Choudhary2022, Cerqueira2023, betenet}. This approach differs fundamentally from ML models trained to predict $\Tc$ directly~\cite{Stanev2018,Hamidieh2018,Meredig2018,Matsumoto2019,Zeng2019,Ishikawa2019,Xie2019,Le2020,Li2020,Dan2020,Hutcheon2020,Roter2020,Konno2021,Shipley2021,Pereti2023}: Since the Eliashberg function captures the full frequency-dependent distribution of the electron-phonon interactions, the effective ``pairing glue,'' it provides deeper physical insight than a single $\Tc$ value and naturally treats superconducting and non-superconducting materials on equal footing.   

\begin{figure*}
    \includegraphics[width=0.8\textwidth]{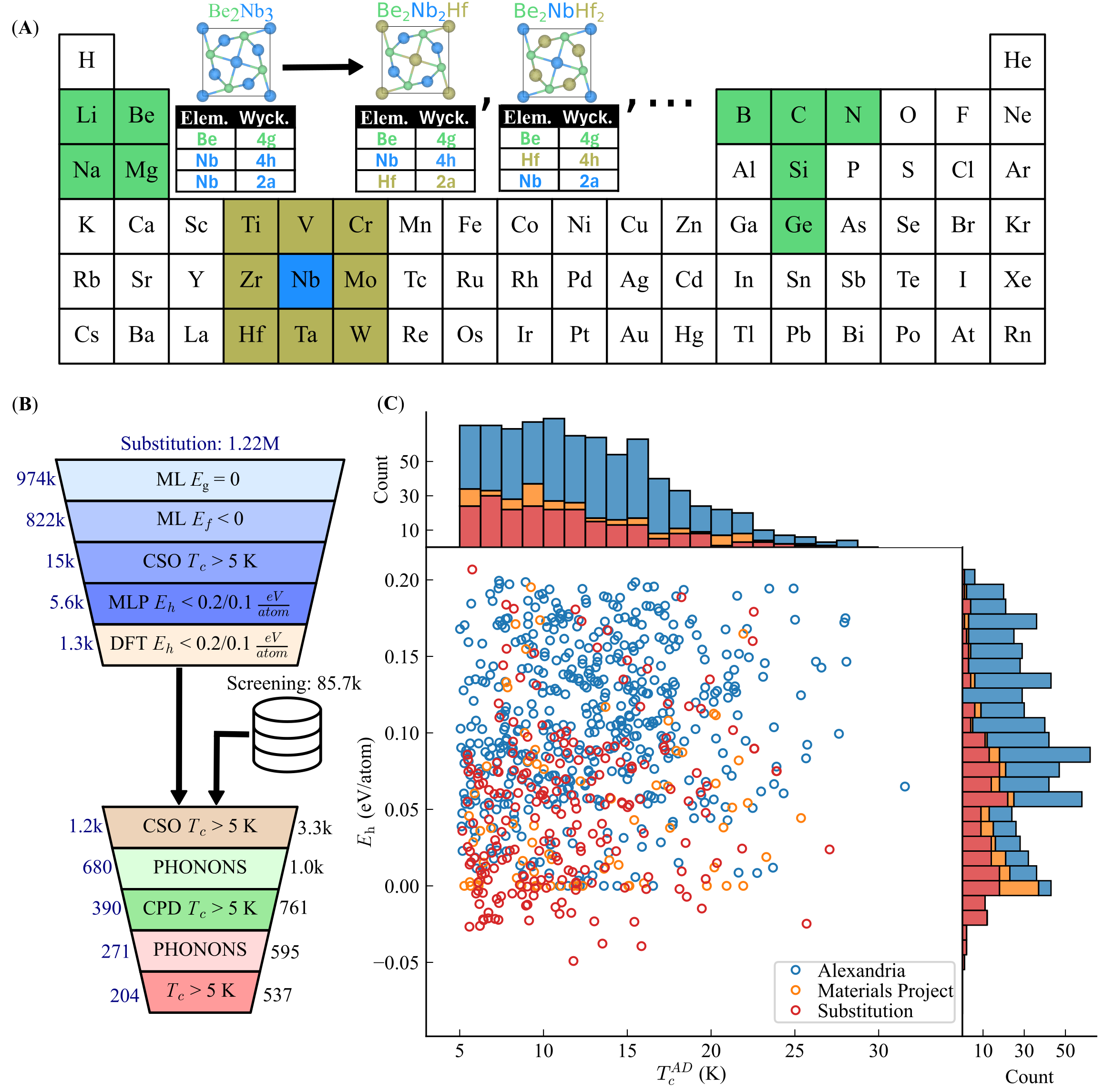}
    \caption{\textbf{Illustration and result overview of the AI-accelerated superconductor discovery workflow.}
    (\textbf{A}) Illustration of the process used to generate structures by partial Wyckoff site substitution (red points in panel \textbf{C}). In a first substitution search, the Materials Project was queried for metals containing the elements highlighted in green. For a given compound, the green element is fixed, and the remaining elements are substituted with neighboring elements on the periodic table. An example is shown for Be$_2$Nb$_3$. In a second substitution search, we repeat the process but query all experimentally stable metals from the Materials Project not containing elements shown in green and no longer fix an element. (\textbf{B}) Workflow for screening materials generated by elemental substitution and queried from the Materials Project and Alexandria database (orange and blue points in panel \textbf{C}). (\textbf{C})~Energy above the convex hull versus DFT-computed $\Tc$ (i.e., via Allen-Dynes equation, $\TcAD$) for all materials that made it to the final stage of the screening process. The histograms show the distribution of material properties.}
    \label{overview}
\end{figure*}

Here we introduce two ML models to predict the Eliashberg spectral function, leveraging a Bootstrapped Ensemble of Equivariant Graph Neural Networks (BEE-NET) architecture. The first model relies exclusively on crystal structure data, making it computationally inexpensive and ideal for large-scale database screening. The second model achieves improved accuracy by additionally incorporating phonon density of states information. After applying the Allen-Dynes equation~\cite{Allen-Dynes1975} (i.e., $\Tc\approx\TcAD$), we achieve a mean absolute error (MAE) as low as 0.9 K in the superconducting transition temperature, enabling rapid screening for new superconductors. By learning $\alpha^2F(\omega)$, our models become capable of training on superconducting and \textit{non}-superconducting materials on equal footing, allowing the reliable ML identification of non-superconductors ($\Tc\leq\SI{5}{K}$) with a true negative rate as high as 0.994. Building on this, we construct a high-throughput virtual screening (HTVS)~\cite{wihtvs} pipeline that integrates ML predictions with first-principles simulations to select for stability, metallicity, and a high pairing strength, winnowing a pool of over 1.3 million materials down to 741 stable superconductors, 69 of which have a predicted $\Tc \geq 20$~K, as illustrated in Figure~\ref{overview}. %
From this set of 741 new superconductors, we present the successful experimental synthesis and characterization for two newly identified superconductors, Be$_2$HfNb$_2$ and Be$_2$HfNb. %

\section{Results}\label{Sec:Results}

\subsection{\boldmath Predicting the Eliashberg spectral function \texorpdfstring{$\alpha^2F(\omega)$}{a2F}}

We trained two variants of BEE-NET to allow for maximally efficient screening: 
The crystal structure only (CSO) variant takes exclusively the crystal structure as input.
In contrast, the coarse phonon density of states (CPD) variant complements the structural input by the coarse phonon density of states (PhDOS). 
During the training process, we use the database constructed by Cerqueira \textit{et al.}~\cite{Cerqueira2023_database}, which consists of 7,000 consistently DFT-computed $\atf$.
We divide this dataset into an 80/20 split for training and testing.
The $\atf$ were binned and smoothed as described in the Methods section, in line with our previous work~\cite{betenet}, and the models were trained to predict the smoothed $\atf$.

We developed in total six models by training the CSO and CPD variants using mean squared error (MSE), weighted mean squared error (WMSE), and the earth mover's distance (EMD)~\cite{EMD} as loss functions (see Supplemental Note~1 in the Supplemental Material for details). %
Figure~\ref{test_results} presents the $\Tc$ predictions for the CSO and CPD variants across the three loss functions, while Table~\ref{tab:model_comparison} summarizes the corresponding regression metrics.
The MSE loss yielded the least accurate predictions for the coupling constant $\lambda$ and transition temperature $\Tc$, although its performance for the frequency moments was comparable. While the WMSE loss function led to significant improvements, the EMD loss function achieved the best overall performance, reducing the MAE by more than 20\% relative to the models trained with MSE loss.

\begin{figure}
    \centering
    \includegraphics[width=\linewidth]{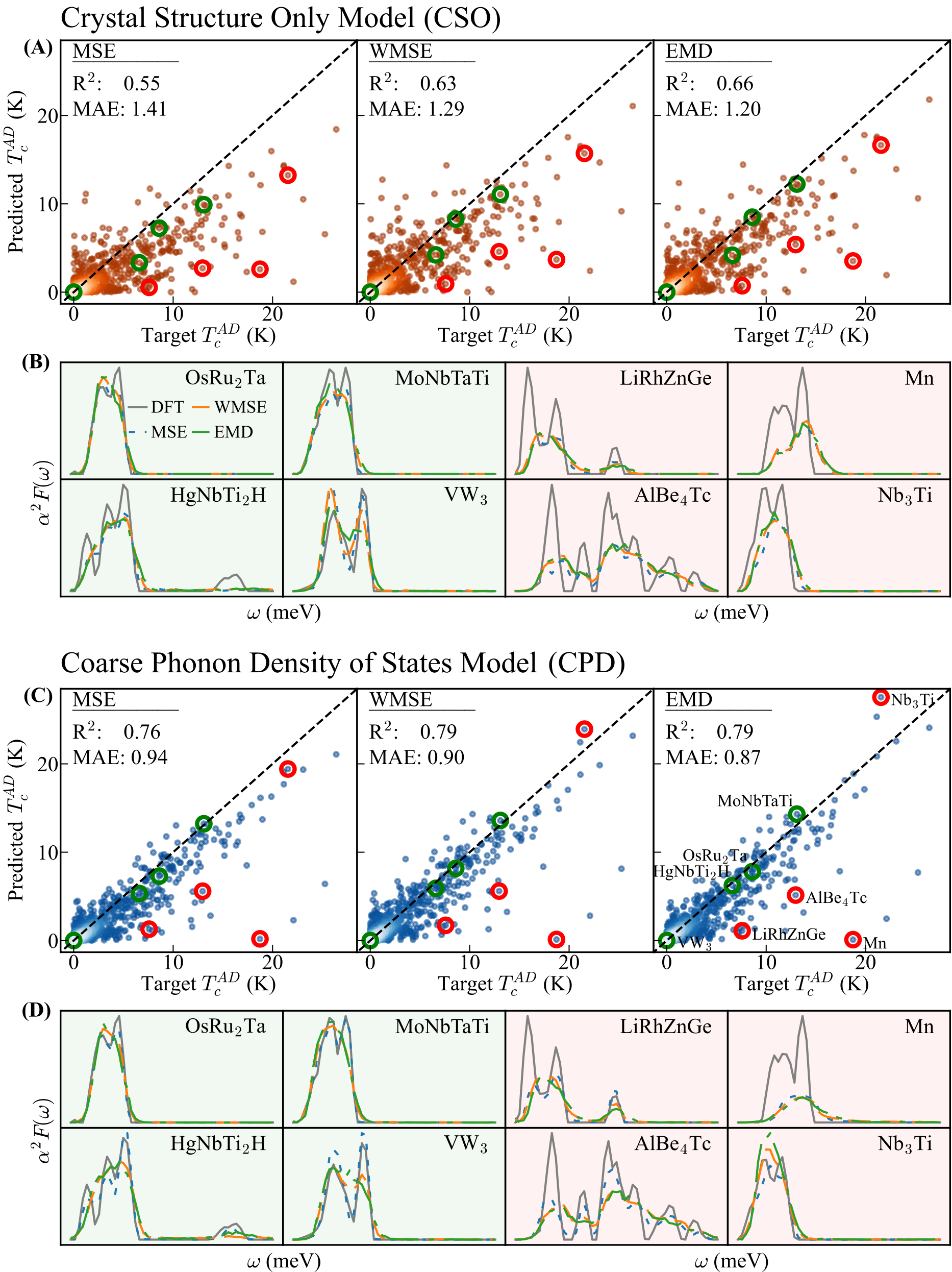}
    \caption{\textbf{Analysis of the test predictions obtained by different model variants and loss functions.}
    (\textbf{A}, \textbf{C}) Comparison of Allen-Dynes $\TcAD$ as obtained from the model-predicted Eliashberg function $\atf$ versus the full DFT result ('target') for the different model variants trained employing either the mean squared error (MSE), weighted mean squared error (WMSE), and the earth movers distance (EMD) loss function. The green and red circles signify representative examples with prediction accuracy in the upper and lower quartile, respectively. The upper and lower quartiles are defined according to the CPD model trained with the EMD loss functions. (\textbf{B}, \textbf{D}) Direct comparison of model-predicted versus DFT-simulated $\atf$ for the highlighted representative materials. Green shading represents the upper quartile predictions, while red shading represents the lower quartile.}
    \label{test_results}
\end{figure}

We find that the models perform exceedingly well in classifying materials with a $\Tc > \SI{5}{K}$. This is quantified by the precision, true positive rate (TPR), and true negative rate (TNR) reported in Table~\ref{tab:model_class}. Precision measures the fraction of predicted superconducting materials ($\Tc > \SI{5}{K}$) that are actually superconducting according to DFT, reflecting the model's reliability in making positive predictions. The TPR, also known as recall, measures the fraction of actual superconducting materials ($\Tc > \SI{5}{K}$) correctly identified by the model, while TNR measures the fraction of materials with $\Tc \leq \SI{5}{K}$ that are correctly classified as such. Although the variants trained with the EMD loss function obtained the best regression metrics, the models trained using the MSE loss function had the highest TNR, with the CSO and CPD variants obtaining $0.98$ and $0.994$, respectively, albeit at the cost of a lower recall.
The remarkably high TNR of the models trained with the MSE loss function ensures a minimum number of redundant calculations when the models are used to screen materials. 
These models were therefore used in the high-throughput screening workflow described in the following section.

\setlength{\tabcolsep}{5pt}
\begin{table}[b]
\centering
\caption{Comparison of R$^2$ and MAE for different loss functions $\mathcal{L}$ across model variants and superconducting properties.}
\label{tab:model_comparison}
\begin{ruledtabular}
\begin{tabular}{cccccccc}
\multirow{2}{*}{\textbf{Prop.}} & \multirow{2}{*}{\textbf{Variant}} & \multicolumn{2}{c}{$\mathcal{L}_\text{MSE}$} & \multicolumn{2}{c}{$\mathcal{L}_\text{WMSE}$} & \multicolumn{2}{c}{$\mathcal{L}_\text{EMD}$} \\ %
                   &                   & \textbf{R$^2$} & \textbf{MAE} & \textbf{R$^2$} & \textbf{MAE} & \textbf{R$^2$} & \textbf{MAE} \\ \hline
\multirow{2}{*}{$\Tc$} & CSO & 0.55 & 1.41 & 0.63 & 1.29 & 0.66 & 1.20 \\ %
                     &    CPD                   & 0.76 & 0.94 & 0.79 & 0.90 & 0.79 & 0.87 \\ \hline
\multirow{2}{*}{$\lambda$} & CSO & 0.57 & 0.109 & 0.63 & 0.102 & 0.65 & 0.096 \\ %
                           &         CPD              & 0.77 & 0.077 & 0.79 & 0.075 & 0.80 & 0.072 \\ \hline
\multirow{2}{*}{$\omega_{\log}$} & CSO & 0.74 & 22.13 & 0.75 & 21.60 & 0.76 & 21.61 \\ %
                                 &           CPD            & 0.86 & 15.30 & 0.86 & 15.92 & 0.83 & 17.33 \\ \hline
\multirow{2}{*}{$\omega_2$} & CSO & 0.86 & 19.67 & 0.87 & 19.31 & 0.87 & 19.12 \\ %
                            &             CPD          & 0.92 & 13.50 & 0.92 & 14.04 & 0.91 & 14.77
\end{tabular}
\end{ruledtabular}
\end{table}

\setlength{\tabcolsep}{2.3pt}
\begin{table}[b]
\centering
\caption{Comparison of precision, TPR, and TNR for different loss functions $\mathcal{L}$ across model variants. TPR and TNR are intrinsic properties of the models when used for classification, meaning they are independent of the class distribution (prevalence) of our test set~\cite{intrinsic} and give an estimate of how the models will perform when screening for novel materials. All values are reported for a classification criteria of $\Tc > \SI{5}{K}$ and the best results are emphasized in bold.}
\label{tab:model_class}
\begin{ruledtabular}
\begin{tabular}{cccccccccc}
\multirow{2}{*}{\textbf{Variant}} & \multicolumn{3}{c}{--- $\mathcal{L}_\text{MSE}$ ---} & \multicolumn{3}{c}{--- $\mathcal{L}_\text{WMSE}$ ---} & \multicolumn{3}{c}{--- $\mathcal{L}_\text{EMD}$ ---} \\ 
& \textbf{Prec.} & \textbf{TPR} & \textbf{TNR} & \textbf{Prec.} & \textbf{TPR} & \textbf{TNR} & \textbf{Prec.} & \textbf{TPR} & \textbf{TNR} \\ \hline
CSO & \textbf{0.84} & 0.47 & \textbf{0.98} & 0.78 &0.61 & 0.96 & 0.80 &\textbf{0.63} & 0.97 \\ %
                            CPD          & \textbf{0.96} &0.63 & \textbf{0.994} & 0.89 &\textbf{0.74} & 0.98 & 0.94 &0.71 & 0.991
\end{tabular}
\end{ruledtabular}
\end{table}

\subsection{\boldmath AI-accelerated screening for high-\texorpdfstring{$\Tc$}{Tc} superconductors}

\begin{figure*}
    \centering
    \includegraphics[width=\textwidth]{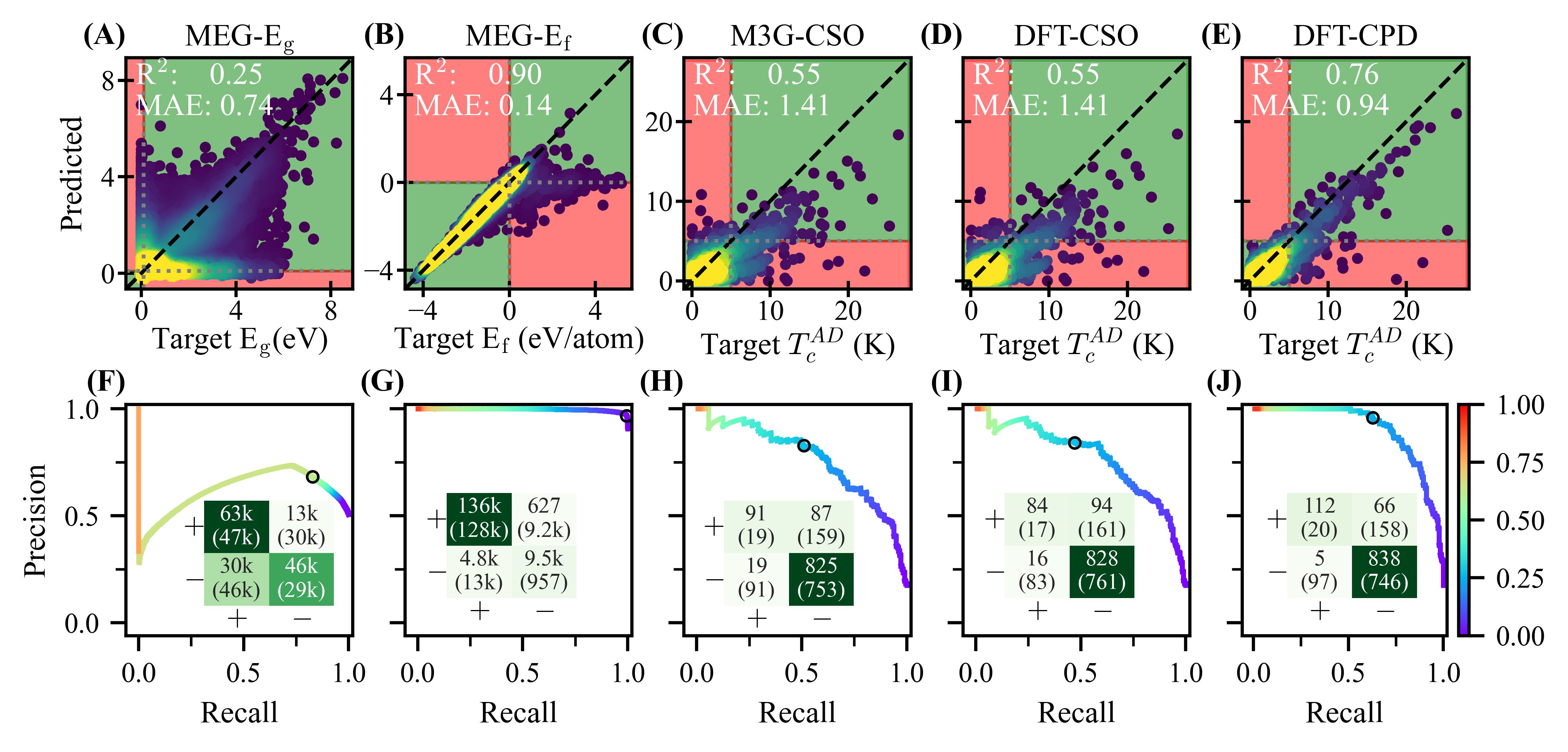}
    \caption{\textbf{Evaluation of the different screening criteria on the test set.} (\textbf{A}--\textbf{E}) Test results of the models used in our screening workflow. The green region represents true positives and true negatives, while the red region shows false positives and false negatives for our screening criteria. (\textbf{F}--\textbf{J}) Precision-recall curves for the models. The marker represents our screening threshold. The inset shows the confusion matrix for this threshold, with the number in parentheses showing the number of materials if a random classifier was used instead.}
    \label{test_pr}
\end{figure*}

To identify promising superconducting candidates, we applied our AI-accelerated screening workflow to two distinct materials sets: known metals queried from materials databases, and novel materials generated through partial Wyckoff site substitution of the occupied sites.

For the first strategy, we searched the Alexandria~\cite{alex1,alex2} and Materials Project~\cite{Jain2013} databases for metals containing up to 12 atoms in the unit cell and an energy above the convex hull ($\Eh$) below 200~meV/atom, yielding a dataset that consisted of 85.7k metals (Figure~\ref{overview}B).

Moreover, to explore superconducting candidates \textit{beyond} known materials, in the second strategy, we generated two sets of novel candidates following the elemental substitution scheme summarized in Figure~\ref{overview}A. The first set of generated candidates
was derived from parent structures obtained by by querying Materials Project~\cite{Jain2013} for experimentally known metals with no more than 12 atoms in the unit cell and $\Eh < 200$~meV/atom.
The parent structures were required to include at least one of the following light elements: Li, Be, B, C, N, Na, Mg, Si, or Ge.
Wyckoff sites in the parent structure not occupied by these elements were substituted iteratively with neighboring elements in the periodic table, resulting in approximately 300k new structures.
A second set of candidates
was constructed from experimentally reported metallic parent structures (from Materials Project) that did not contain the aforementioned elements and met a more stringent thermodynamic stability criterion ($\Eh < 100$~meV/atom). The same Wyckoff substitution protocol was applied to this second set of parent structures, yielding an additional 916k candidates. The combined set of generated structures totaled approximately 1.22 million novel materials.

Across both the queried and generated structures, our goal was to identify thermodynamically and dynamically stable metals with a DFT-calculated $\Tc > \SI{5}{K}$. Thermodynamic stability was defined as $\Eh \leq 100$~meV/atom for the second set of generated candidates and $\Eh \leq 200$~meV/atom for all remaining candidates. Dynamically stable candidates do not exhibit imaginary frequencies in the calculated phonon spectrum. Overall, our high-throughput screening task is formalized by answering the following five questions for each candidate material:
\begin{enumerate}
    \item Is the material a metal? (i.e., $\Eg = 0$) %
    \item Is the material stable against decomposition into its constituent elements? (i.e., $\Ef \leq 0$)
    \item Is the material thermodynamically stable?
    \item Is the material dynamically stable?\newline (i.e., $\omega_{q} \in \mathbb{R}$) %
    \item Does the material exhibit superconductivity above 5~K? (i.e., $\Tc > 5$~K)
\end{enumerate}
\noindent
We address these questions in the workflow displayed in Figure~\ref{overview}B by consecutively applying the ML filters shown in Figure~\ref{test_pr}.

For candidates queried from the databases, the first three questions can be answered directly using the reported formation energies, band gaps, and convex hull energies. In contrast, for the generated structures, direct prediction of target properties would incur a substantial  error, since the structures are not yet relaxed~\cite{Witman2023, Gibson2022}. To address this, we first relaxed the generated candidates using the M3GNET~\cite{m3gnet}, a machine-learning model for interatomic potentials. The formation energy is then predicted using the MEGNET model~\cite{megnet}, removing any material with a positive $\Ef$. Next, the band gap is predicted using MEGNET, and any nonmetals are removed. Subsequently, $\Tc$ is predicted using CSO BEE-NET, removing any material with $\Tc \leq 5$~K. $\Eh$ is computed based on the M3GNET final energy for these predicted superconductors ($\Tc > \SI{5}{K}$), and any material with $\Eh$ above the defined threshold was removed. At this point, 1.22 million candidates have been reduced to 5.6k promising candidate superconductors without a single DFT calculation (Figure~\ref{overview}B), highlighting the efficiency of the applied filters that require only seconds of computation per material. This makes it feasible to screen tens of millions of candidate materials per month.

The remaining 5,600 candidates were then structurally optimized via DFT using the Pymatgen MPRelaxMetalSet~\cite{wang2021} parameters. At this stage, the information on %
the database-derived and substitution-generated structures is equivalent, and addressing questions 4 and 5 for both data sets follows the same approach. For each relaxed structure, $\Tc$ was re-evaluated using the CSO BEE-NET model, and those with $\Tc \leq \SI{5}{K}$ were discarded. To efficiently assess the dynamical stability and to avoid redundant calculations of unstable materials, the phonon spectra are computed exclusively on a coarse $2\times2\times2$ $\bq$-grid. Only metals with purely real phonon frequencies are retained. For these dynamically stable candidates, the phonon density of states (PhDOS, calculated from the coarse $2\times2\times2$ $\bq$-grid) along with the DFT-optimized crystal structure is used as input to the CPD BEE-NET model, which provides a more accurate $\Tc$ prediction, again removing materials with $\Tc \leq \SI{5}{K}$. Finally, $\atf$ is computed with high accuracy from DFT for the remaining candidates, yielding 741 stable metals with $\Tc > \SI{5}{K}$.
This last step required only 866 computationally expensive $\atf$ calculations, resulting in a final precision of 86\%. 

\begin{figure}
    \centering
    \includegraphics[width=0.9\linewidth]{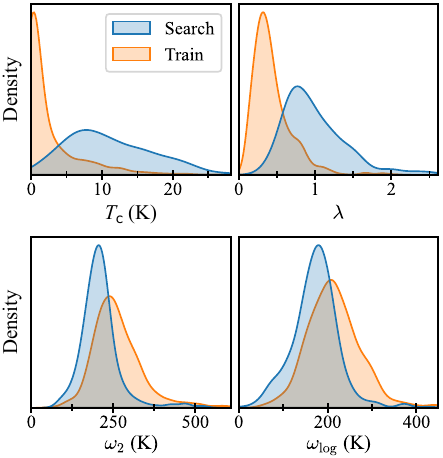}
    \caption{\textbf{Distribution of different superconductivity observables for the training materials compared to the discovered materials.} The distribution of $\Tc$, $\lambda$, $\wtwo$ and $\omega_{log}$ for the training set (orange) and the materials identified by the HTVS workflows (blue).
    The broader $\Tc$ distribution with a significantly higher mean indicates the extrapolation beyond the scope of the training data.}
    \label{dist_plots}
\end{figure}

The final results of this multi-step screening are summarized in Figure~\ref{overview}C, displaying $\Eh$ and $\Tc$ for the discovered materials. A substantial number of materials exhibit simultaneously a low $\Eh$ and a high $\Tc$, which underscores the success of the workflow in identifying stable, high-performance superconductors.
Interestingly, when compared to the training dataset used to develop the models, the screened candidates display a markedly different distribution,
as analyzed in Figure~\ref{dist_plots}.
While the training set is skewed towards low-$\Tc$ materials with a mean $\Tc$ of \SI{2.4}{K}, the screened set shows a broader distribution with a significantly higher mean $\Tc$ of \SI{11}{K}, indicating that the workflow effectively extrapolated beyond the bias of the training data. 

While the final results are already impressive, the overall workflow can be further refined by systematically evaluating the contributions and limitations of each filter. A detailed discussion of each filtration step is provided in the Methods section. Nevertheless, the HTVS workflows introduced here highlight the effectiveness of integrating ML-driven models with physics-informed filtering, enabling the efficient discovery of promising superconductors within an expansive search space.

\subsection{From Theory to Reality: Experimental growth and verification of superconductivity}

\begin{figure*}
    \centering
        \includegraphics[width=0.33\linewidth]{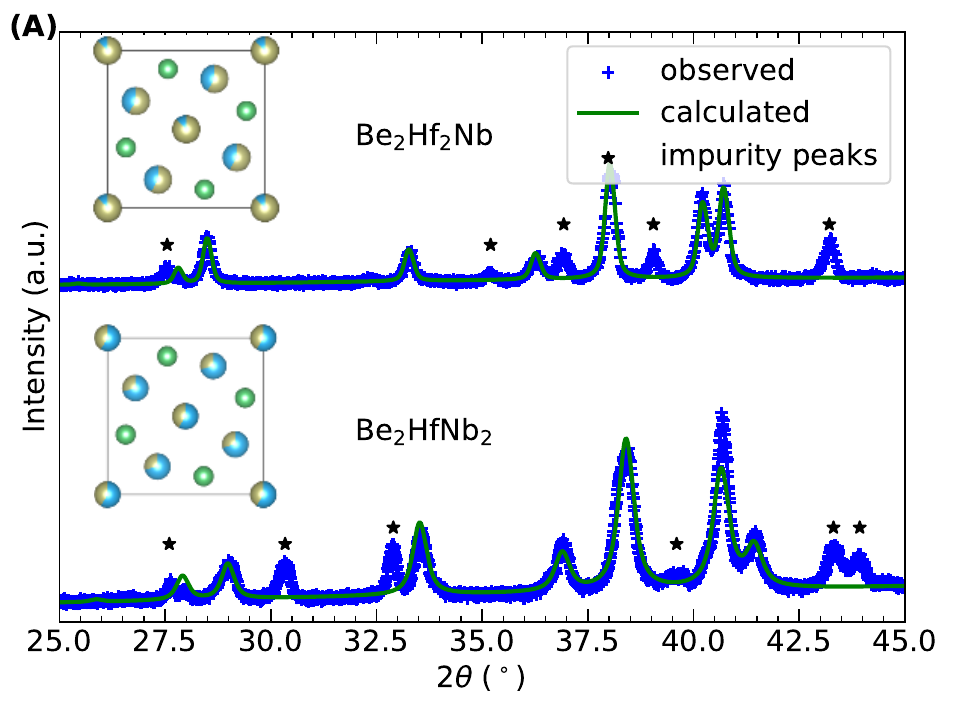}
        \includegraphics[width=0.33\linewidth]{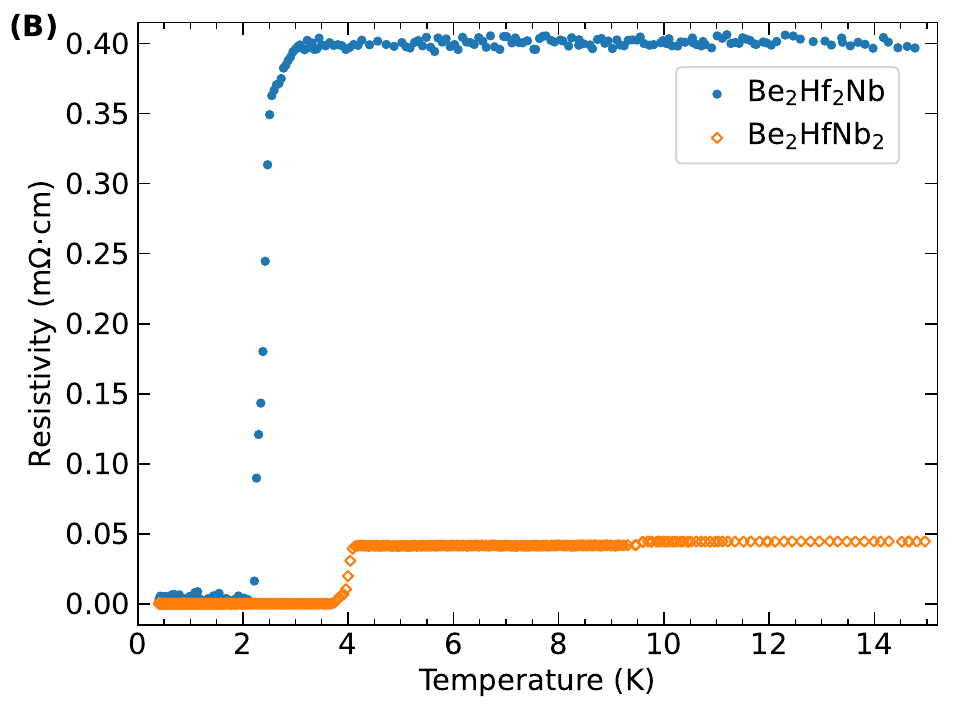}
        \includegraphics[width=0.33\linewidth]{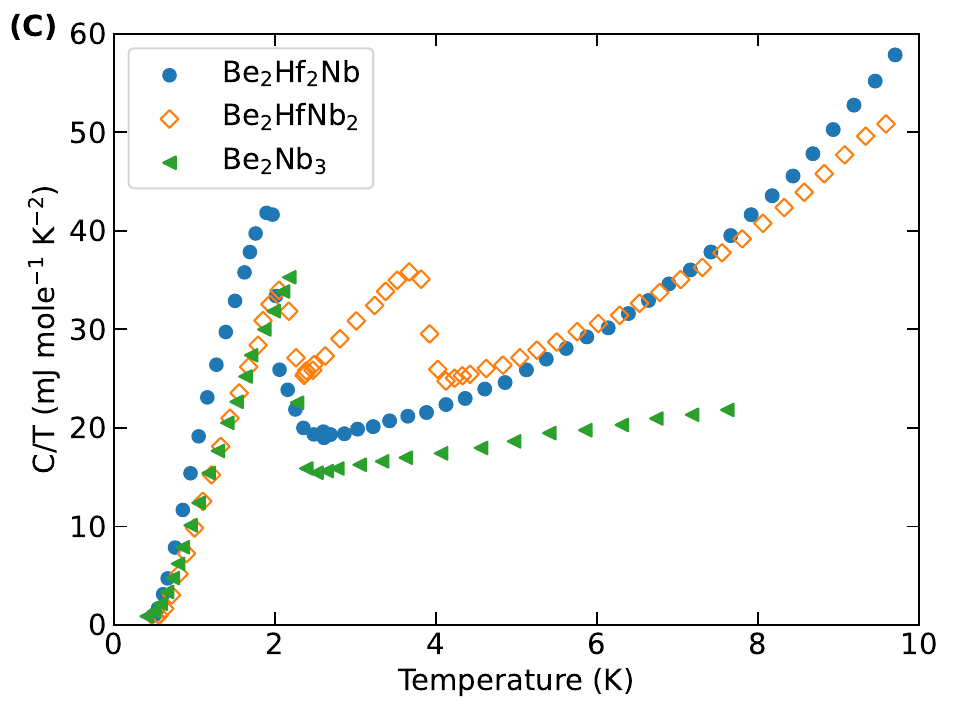}
    \caption{\textbf{Experimental data of Be$_2$Hf$_2$Nb and Be$_2$HfNb$_2$.} 
    \textbf{(A)} The measured and calculated XRD data of Be$_2$Hf$_2$Nb and Be$_2$HfNb$_2$. The calculated intensity is obtained by Rietveld refinements. The insets show the crystal structures and atomic occupancies of Be (green), Hf (brown), and Nb (blue) for each sample [see Fig.~\ref{overview}A for the labeling of the Wyckoff positions (2a, 4h, and 4g)].
    \textbf{(B)} The resistivity data displays the superconducting transitions of Be$_2$Hf$_2$Nb and Be$_2$HfNb$_2$. The slight resistivity drop near 9~K for Be$_2$HfNb$_2$ indicates trace amounts of Nb. 
    \textbf{(C)} Specific heat data.
    \ch{Be2Hf2Nb} and \ch{Be2Nb3} exhibit a single jump, while \ch{Be2HfNb2} shows two distinct jumps indicating two different superconducting phases.
    }
    \label{resistivity}
\end{figure*}

A fundamental challenge in modern HTVS approaches is the sheer volume of predicted promising material candidates, far exceeding what can feasibly be validated experimentally. This imbalance outpaces the conventional collaborative workflow between theory and experiment.
In order to prioritize which materials should be examined experimentally, we narrowed our search to candidates that had a relatively high $\Tc$, low $\Eh$, and a parent compound that was known to be a conventional (electron-phonon) superconductor. 
With this approach, we identified Be$_{2}$Hf$_{2}$Nb ($\Eh^{\mathrm{DFT}} = 94\,\mathrm{\frac{meV}{atom}}$, $\Tc^{\mathrm{DFT}}=14.9$ K) and Be$_{2}$HfNb$_{2}$ ($\Eh^{\mathrm{DFT}} = 62\,\mathrm{\frac{meV}{atom}}$, $\Tc^{\mathrm{DFT}}=7.5$ K). 
These materials were generated by substituting either the 4h or 2a Nb Wyckoff sites of Be$_2$Nb$_3$ with Hf (see Figure~\ref{overview}A).
\ch{Be2Nb3} is known to become superconducting at \SI{2.3}{K}~\cite{Havinga1969}.
To the best of our knowledge, neither of the Hf-containing compositions has been previously studied experimentally or computationally.

To obtain a more accurate prediction of $\Tc$, we recompute $\atf$ using DFT, increasing the $\bk$-point density by a factor of $\sqrt{2}$. We selected a Coulomb pseudopotential of $\mu^{\ast}=0.21$, which accurately describes $\Tc$ for the parent compound Be$_2$Nb$_3$.  
This resulted in $\Tc^{\mathrm{DFT}} = 9.3$~K and 5.1 K for Be$_{2}$Hf$_{2}$Nb and Be$_{2}$HfNb$_{2}$, respectively. 
Given the similar ionic radius of Nb and Hf, Be$_{2}$Hf$_{2}$Nb may have a propensity for disorder.
Therefore, we performed explicit calculations of $\atf$ for all possible disordered states of Nb and Hf, which uncovered that all dynamically stable structures are superconducting as well, with the lower bound of $\Tc$ being 4.2~K.

We synthesized \ch{Be3Nb2}, \ch{Be2Hf2Nb}, and \ch{Be2HfNb2} as described in the Methods section.
We analyzed the x-ray data using LeBail and Rietveld analysis with the GSAS-II software package~\cite{GSASii}.
Details of the x-ray data analysis are provided in the Supplemental Material.
Here, we summarize the main points:
For \ch{Be2Nb3}, our x-ray data is well described by the known \ch{Be2Nb3} structure together with a small amount of \ch{Be2Nb}.
The presence of \ch{Be2Nb} as a secondary phase is not unexpected given the known binary phase diagram for Nb-Be, which shows that \ch{Be2Nb3} melts incongruently.

Initial analysis of the diffraction patterns for \ch{Be2Hf2Nb} and \ch{Be2HfNb2} showed the presence of impurity peaks and some disagreement in the intensity of the peaks corresponding to the predicted structure.
However, as discussed in the Supplemental Material, adding Hf (which is larger than Nb) to the composition produces a clear expansion of the lattice, demonstrating that Hf was successfully introduced into the structure.

Due to the chemical similarity of Hf and Nb, we considered the possibility that Hf and Nb could substitute randomly for each other.
The 4h (vertex) site is coordinated by two nearest-neighbor Be atoms at a distance of \SI{2.58}{\angstrom} and four more Be atoms at a slightly larger distance of \SI{2.66}{\angstrom}.
The 2a (innner) site is coordinated by four Be atoms at a distance of \SI{2.81}{\angstrom}.
The numbers listed here are those for the refined \ch{Be2Hf2Nb} data.
Since there is more ``room'' on the 2a sites, we might expect the larger Hf atoms to preferentially occupy the 2a sites.
This intuition is born out by a more detailed quantitative analysis of the diffraction data summarized in the Supplemental Material (Supplemental Note~2).
We find that for \ch{Be2Hf2Nb}, the Hf:Nb ratios are 87.5\%:12.5\% and 56.25\%:43.75\% for the 2a and 4h sites, respectively.
For \ch{Be2HfNb2}, the Hf:Nb ratios are 41.5\%:58.5\% and 29.25\%:70.75\% for the 2a and 4h sites, respectively.

With the presence of Nb/Hf disorder accounted for, Rietveld analysis produces a reasonable fit to the majority of the peaks for both \ch{Be2Hf2Nb} and \ch{Be2HfNb2} (Figure~\ref{resistivity}A).
The dominant impurity phase appears to be different for the two compositions.
Efforts to match the impurity peaks to known binary or predicted ternary phases in the Nb-Hf-Be phase space have been unsuccessful.
Additional details of the Rietveld fits and the impurity phases that were evaluated are provided in the Supplemental Material.

Low-temperature transport measurements show that Be$_{2}$Hf$_{2}$Nb and Be$_{2}$HfNb$_{2}$ both exhibit superconductivity, with onset $\Tc$ values of 3.18~K and 4.24~K, respectively (Figure~\ref{resistivity}B).
For both samples, the residual resistivity ratio (RRR) is about 1, consistent with the presence of substantial disorder.
Disorder arising from defects can be ruled out, as the RRR does not improve after annealing.

Specific heat data on the three compounds is shown in Figure~\ref{resistivity}C.
For \ch{Be2Nb3}, we find $\Delta C/(\gamma T_c) = 1.42$, which is consistent with bulk superconductivity.
For \ch{Be2Hf2Nb}, we %
find a single transition with $\Delta C/(\gamma T_c) = 1.44$, consistent with bulk superconductivity and an onset close to (but slightly below) that observed in the electrical resistance measurement.
Together with the evidence for lattice expansion from the x-ray data, the low-temperature measurements demonstrate that we successfully synthesized a sample with composition \ch{Be2Hf2Nb} that is fully superconducting below $\sim \SI{2}{K}$.
In the case of \ch{Be2HfNb2}, we identify two specific-heat jumps that are roughly comparable in size, suggesting a multi-phase sample with superconducting secondary phase.

\section{Discussion}\label{Sec:Discussion}
Superconductors have the potential to transform power transmission and magnetic technology, yet their discovery remains largely serendipitous. High-throughput virtual screening (HTVS) is limited by the computational cost of estimating the critical temperature, and past machine learning models often lacked the accuracy needed for effective superconductor discovery. Moreover, most HTVS approaches predict superconducting properties in a single step, missing valuable insights gained through progressive filtering.

To address these challenges, we developed BEE-NET, a deep-learning model with state-of-the-art accuracy in predicting superconducting properties. BEE-NET is a bootstrapped ensemble of equivariant graph convolutional neural networks trained to predict the Eliashberg spectral function accurately. By employing a variant of the Earth Mover’s Distance as an alternative to the MSE loss function, we significantly improved the generalizability of spectral function predictions. We integrated BEE-NET with additional machine learning models and density functional theory to construct a comprehensive AI-accelerated workflow for superconductor discovery. This workflow enabled high-throughput virtual screening with 92\% precision for existing materials and 76\% for generated candidates, ultimately identifying 741 stable superconductors with an overall precision of 86\%. Finally, we experimentally synthesized two predicted materials and confirmed their superconducting behavior.

The flexibility of our approach offers the potential for further improvements.
We speculate that adding a DFT confirmation that a candidate is a metal would likely further improve the precision of the generated materials. The computational cost of screening could be further reduced by developing and integrating a machine-learned surrogate for predicting dynamic stability. Furthermore, because our models have been designed to predict $\atf$ up to a maximum frequency of 100 meV and were trained on ambient pressure data, they are unlikely to accurately predict high-temperature superconductors similar to hydrides. This important goal will require further expansion of the training set. Intriguingly, the discovery of a new 3D ductile material with $\Tc$ above 30~K is possible within our current workflow. Such a system would already have revolutionary implications for applied superconductivity in magnets and other applications.

These results underscore the power of integrating machine learning, computational methods, and experimental techniques to accelerate materials discovery. The presented AI-driven HTVS workflow not only successfully identified experimentally synthesizable superconductors, but also established a scalable, systematic approach for uncovering novel materials. This work moves machine learning beyond theoretical promise, demonstrating its practical role in revolutionizing materials discovery. With continued advancements, such frameworks could drive the discovery of next-generation superconductors, enabling energy-efficient power grids, lossless electronics, and magnetically levitated transport, key innovations for a sustainable, high-tech future.

\section{Methods}\label{Sec:Methods}

\subsection{Density functional theory calculations, data preparation, and model training}

For $\Eh$ calculations, we use the \textit{Vienna Ab initio Simulation Package} (VASP) \cite{Kresse1996, Kresse1996b, Kresse1999} with parameters defined in the Pymatgen MPRelaxMetalSet~\cite{wang2021}. We compute $\Tc$ by utilizing Quantum Espresso~\cite{qe1,qe2,qe3} with the PBEsol exchange-correlation functional. We follow the methodology outlined in our previous work~\cite{betenet} to generate the commensurate $\bk$- and $\bq$-point meshes for electron-phonon calculations and use the same DFT parameters.

For training variants of the BEE-NET CSO and CPD models, we use the computed $\atf$ dataset from Cerqueira \textit{et al.}~\cite{Cerqueira2023_database}. The $\atf$ values are binned and smoothed using the Savitzky-Golay filter, following the same procedure as our previous work~\cite{betenet}. The dataset is first split into 80\% training and 20\% test sets. Within the training set, we generate a bootstrapped dataset for training 100 ensembles of the CSO and CPD models, where each bootstrapped training set contains approximately 62\% unique data points from the original training split, while the remaining 38\% of the unique data points are reserved for validation.

We train three variants of both the CSO and CPD models (six models in total) using MSE, WMSE, and EMD loss functions, with the AdamW optimizer implemented in PyTorch~\cite{PyTorch}. We use a fixed learning rate of 0.005 and a weight decay of $1 \times 10^{-7}$. The model architecture includes a cutoff radius of 4~$\AA$, an embedded feature length of 64, an irreducible multiplicity of 32, two point-wise convolution layers, and 10 radial basis functions with the radial network consisting of a single layer with 100 neurons in the head. Further details on the Euclidean neural network (e3NN) architecture can be found in Refs.~\onlinecite{e3nn_1, e3nn_2, e3nn_3}.

\subsection{Experimental synthesis}
\label{sec:synth_methods}
In order to close the theoretical-experimental loop in our search for new superconductors in the Be$_{2}$Nb$_{3}$ system, we prepared Be$_{2}$Nb$_{3}$ and the Hf-substituted variants Be$_{2}$Hf$_{2}$Nb and Be$_{2}$HfNb$_{2}$.
Preparation was done by arc-melting together Nb, Be, and\textemdash for the Hf-doped samples\textemdash Hf.
Since Nb has a 10 times larger atomic mass than Be (92.906 vs.\ 9.01218), and similarly Hf has a $\sim$20 times larger atomic mass (178.49) than Be,  half gram buttons had typically only about 40 mg or less Be to reach the desired stoichiometric end point.
Samples were melted three times and flipped over between meltings.
Since each melting vaporized a significant amount of Be, the melting process started with an excess of approximately five times the stoichiometric amount of Be, which by sequential careful melting ended in the required Be amount within a few~mg.

\subsection{Details of the screening workflow and filtering}
Figure~\ref{test_pr} shows the trade-off between precision and recall for each ML filter in the workflow. 
`Precision' estimates how many of the materials retrieved by the ML model are relevant, while `recall' estimates how many of the relevant materials are retrieved. 
Using these metrics allows an estimate of the datasets after the ML filter is applied. 

The generated materials dataset originally consisted of 1.22 million candidates, which was reduced to 974k after the application of the $\Eg$ filter shown in Figures~\ref{test_pr}(A, F).
Despite the relatively poor predictions of this ML model (R$^2$ = 0.25), it still obtains a reasonable precision and recall of 0.681 and 0.829, respectively.  

Application of the $\Ef$ filter, shown in Figures~\ref{test_pr}(B, G), further reduced the dataset to 822k candidates. The models' precision and recall are 0.966 and 0.995, respectively. 
While we assume that the number of overlooked structures is exceptionally low, the number of thermodynamically stable candidates overlooked is expected to be far lower.
This is because although the model slightly overestimated $\Ef$ for the overlooked candidates, a formation energy close to zero is still high and the subsequent $\Eh$ filter would have likely removed these candidates. 

Subsequently, the CSO variant of BEE-NET [Figures~\ref{test_pr}(C, H)] was applied to predict $\atf$, resulting in 15k materials with $\Tc > \SI{5}{K}$.
Such a sizeable reduction is not surprising, as high-$\Tc$ superconductors are exceedingly rare.
Despite being trained on DFT-relaxed structures, the model performs well on structures that are instead relaxed using M3GNET, incurring only a 0.04~K increase in the test MAE. 
With a screening threshold of 5~K, the model presents a recall of 0.511 and a precision of 0.827.  
This step reduces the number of candidates to 15k, of which an estimated 7.5k are likely to exhibit a $\Tc$ above 5~K based on the model’s precision.

Thermodynamic stability is predicted by computing $\Eh$ based on the Materials Project phase diagram and M3GNET final energy.
An $\Eh$ filter of 200~$\mrm{\frac{meV}{atom}}$ further reduced the number of candidates to 5.6k. 
Reducing 1.22 million candidates to 5,600 without a single DFT calculation highlights the efficiency of the applied filters and requires only seconds of computation per material.

To confirm $\Eh$, the remaining materials are relaxed using DFT with the Pymatgen MPRelaxMetalSet parameters~\cite{wang2021}.
This further reduces the number of candidate structures down to 1,300 --- a sizable drop, considering that $\Eh$ was previously predicted by the M3GNET model. 
However, the $\Eh$ predictions were not evaluated on a test set as the prediction of $\Ef$ was deemed sufficient, and computing $\Eh$ for all materials on the Materials Project based on the final M3GNET energy is a burdensome task. 
A retrospective analysis (Supplementary note~2) revealed that M3GNET systematically underpredicts $\Eh$ by approximately 140~$\mathrm{\frac{meV}{atom}}$, which, while lowering precision, acted as a more lenient screening threshold, favoring higher recall over precision. Correcting this underestimation would have reduced the recall to, at most, 52\%, excluding a substantial number of viable candidates.

Applying the CSO model to the generated, DFT-relaxed structures [Figure~\ref{test_pr}(D, I)] removed an additional 67 candidates. 
This small drop is expected, as the predictions on DFT-relaxed and M3GNET-relaxed structures are highly correlated.
In contrast, there is a stark drop from 85.7k to just 3.3k structures for the materials screened from databases. Again, this is an expected drop based on the same rationale as the M3GNET structures.

After application of the PhDOS filter, 1k and 680 candidates remain for the screened and generated datasets, respectively. The CPD model is applied further [Figure~\ref{test_pr}(E, J)], reducing the number of candidates to 761 and 390, respectively. 
Finally, $\atf$ is computed, and of those materials for which the calculations converged successfully, 537 and 204 exhibited a DFT-computed $\Tc$ greater than 5~K.

\section{Acknowledgments}
This work was funded by the U.S. National Science Foundation, Division of Materials Research, under Contract No.\ NSF-DMR-2118718. A.C.H.\ and R.G.H.\ acknowledge additional support from the National Science Foundation under award PHY-1549132 (Center for Bright Beams). 
P.M.D. acknowledges support from the Laboratory Directed Research and Development
Program of Oak Ridge National Laboratory, managed by UT-Battelle, LLC, for the US Department of Energy.
Computational resources were provided by the University of Florida Research Computing Center.
B.G., R.G.H., and P.J.H.\ acknowledge support of this work by the ``AI and Complex Computational Research'' funding framework of the University of Florida. 

\section{Author Contributions}
J.B.G. and R.G.H. conceived the project. A.C.H. performed DFT calculations for the training database generation. J.B.G. performed all model implementation and training and DFT calculations for the screened materials. P.M.D., B.G., and R.G.H. provided context into the theory of superconductivity and conceptualized alternative loss functions. G.R.S. and J.S.K. synthesized the samples and measured superconductivity. J.J.H. and Z.L. ran the XRD and performed the structural analysis. B.G., P.J.H., and R.G.H. supervised the research. J.B.G., P.M.D., B.G., P.J.H., Z.L., G.R.S., and R.G.H. wrote the manuscript. All authors contributed to revising and editing the manuscript.

\end{document}